\documentclass[review]{elsarticle}
\usepackage{amsmath}
\usepackage{cases}

\journal{Optics Communications}

\begin{document}

\begin{frontmatter}

\title{Structured Illumination Microscopy Based on Fiber Devices}


\author[mymainaddress,mysecondaryaddress]{Shiming Hu}
\author[mymainaddress]{Wenwen Liu}
\author[mymainaddress,mysecondaryaddress]{Junyao Jie}
\author[mymainaddress,mysecondaryaddress]{Yizheng Huang}
\author[mymainaddress]{Qingquan Wei}
\author[mymainaddress,mysecondaryaddress]{Manqing Tan}

\author[mymainaddress,mysecondaryaddress]{Yude Yu\corref{mycorrespondingauthor}}
\cortext[mycorrespondingauthor]{Corresponding author}
\ead{yudeyu@semi.ac.cn}

\address[mymainaddress]{State Key Laboratory of Integrated Optoelectronics, Institute of Semiconductors, Chinese Academy of Sciences, Beijing 100083, China}
\address[mysecondaryaddress]{College of Materials Science and Opto-Electronic Technology, University of Chinese Academy of Sciences, Beijing 100049, China}

\begin{abstract}
We present a simple and compact approach of structured illumination microscopy by using three $2\times2$ fiber couplers and one $1\times4$ MEMS optics switch.
One uniform and three fringe illumination patterns were produced by placing seven output fiber tips at the conjugate Fourier plane of the illumination path.
Stable and relatively high-speed illumination switching was achieved by the optics switch.
Super-resolution and optical sectioned information was reconstructed from 4-frame data by using algorithms based on a joint Richardson-Lucy deconvolution method and a Hilbert transform method.
By directly removing the out-focus components from the raw images provides better imaging results.
\end{abstract}

\begin{keyword}
Imaging Systems \sep Microscopy \sep Structured Illumination \sep Superresolution \sep Optical Section
\end{keyword}

\end{frontmatter}


\section{Introduction}
Structured illumination microscopy (SIM) has been an efficient way to enhance the resolution of the fluorescent microscopy \cite{Gustafsson2000}.
Compared to confocal microscopy, SIM has the advantages of being a wide-field imaging system, and has the same optical section capability as well \cite{neil1997method,gustafsson2008three}.
To achieve both super-resolution (SR) and optical sectioning (OS), SIM requires raw SI images with three or five phase-shifts at three orientations.
To fulfill this requirements, structured illumination (SI) methods with phase shift have been achieved by using gratings \cite{Wang2011}, SLMs \cite{Chang2009,dan2013dmd}, integrated optics chips \cite{liu2016structured,helle2019structured}, galvanometers \cite{liu2019three}, and other devices \cite{rodriguez2008axial,Wei2014}.

Actually, the desire of phase shift makes it complex and expensive to build a SIM system, which needs those devices to modulate the phase accurately.
Recently, several SR algorithms that require only four raw images (one wide-field and three SI images) have been proposed \cite{orieux2011bayesian,Dong2015,Lal2016a,Meiniel2017}.
Meanwhile, it is also possible to achieve OS from non-phase-shift raw images \cite{hoffman2017single,jost2015optical,soubies2018computational}.
On one hand, those methods increase the frame rate by reducing the number of raw images; on the other hand, they do not require any phase shift of the SI pattern, which could be used to simplify the optical setup.
In our previous work \cite{hu2019compact}, we placed a $2 \times 2$ fiber coupler's output tips at the conjugate Fourier plane of the illumination path to produce fringe patterns at the sample plane.
By rotating the fiber mount, SI images at multiple orientations ($\geq 3$) without any phase shift were captured and SR result was reconstructed by a joint Richardson-Lucy (jRL) deconvolution method \cite{Strohl2015}.
Although this method is simple and low-cost, both of the quality of the SI pattern and the imaging speed of system are limited by the mechanical rotation of the fiber tips.

In this paper, we proposed an improved SIM system using three $2 \times 2$ fiber couplers and a $1 \times 4$ MEMS optics switch with no moving part.
We modified a commercial microscope (IX-83, Olympus) by placing seven fiber tips at the conjugate Fourier plane of the illumination path, which could provide three SI and one uniform illumination at the sample plane.
The use of the optics switch significantly increased the speed of SI pattern generation and the stability of the system, comparing to our previous work \cite{hu2019compact}.
Both SR and OS images were reconstruct by using a jRL deconvolution method and a single-slice optical section algorithm based on Hilbert transform.
In addition, we presented two different strategies to merge the SR and OS components, and it shows that by directly removing out-focus information from each raw images will yield a better imaging result.

\section{Method}
\begin{figure}
\centering
\includegraphics{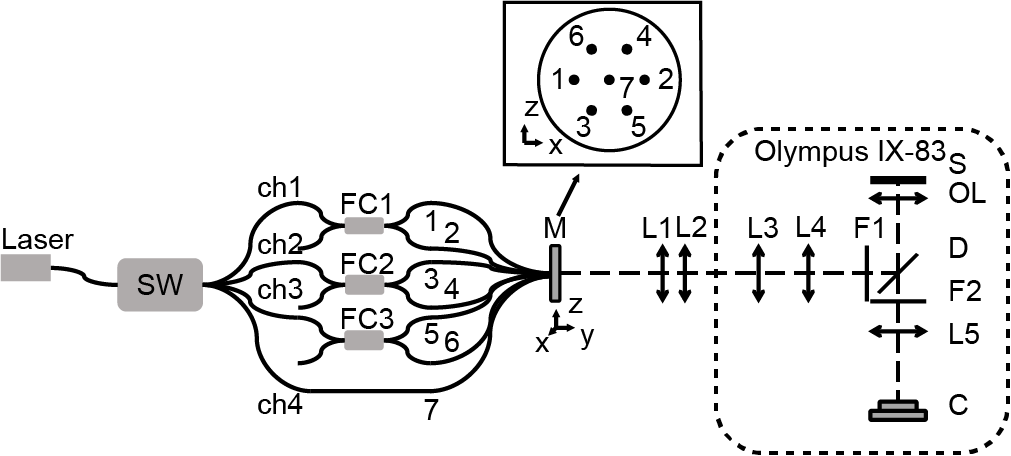}
\caption{An schematic diagram of our SIM system based on an Olympus IX83 microscope. SW: a $1 \times 4$ MEMS optics switch; FC1-FC3: three $2 \times 2$ fiber couplers; M: a 3D printed fiber mount;  L1-L5: lenses; OL: objective lens; F1: an excitation filter; F2: an emission filter; D: a diachronic mirror; S: sample plane. C: a digital camera. The layout of the fiber tips' position (1-7) mounted on M is shown in the upper panel.}
\label{fig: setup}
\end{figure}

Fig. \ref{fig: setup} shows a schematic diagram of our optical setup.
A 532 nm laser source (LC-651A Laser Combiner, ColdSpring) is directly coupled into a $1\times4$ MEMS optics switch (Ziguan, Sichuan China), SW.
Three of switch's output channels (ch1-ch3) are connected to three $2\times2$ single mode fused fiber optic couplers (TN532R5F2, THORLABS), FC1-FC3, with 50:50 coupling ratio, whose output tips are mounted on a 3D printed fiber mount, M, in concentric arrangement. And the remained channel (ch4) is directly mounted on the center of the fiber mount, while the fiber mount is placed at the conjugate Fourier plane of the illumination path. The layout of fiber tips' position is shown in the upper panel of the Fig. \ref{fig: setup}.
All the fiber tips are placed in parallel and they directly outputted the light from 2.5 $\mu m$ cores which could be considered as point light sources.
Two lenses (L1-L2) are used to collect and projected the light into the back aperture of the microscope.
The z-scan moving stage we used could provided 0.01 $\mu m$ resolution in focusing control with maximum speed of 5 mm/sec.
A sCMOS camera (ORCA-Flash 4.0 V2, HAMAMATSU) is used to capture the fluorescent signal.
The tube lens of the system, L5, is a 180 mm lens.

In this setup, each of the fiber couplers could split the incident light into two beams which could produce an interference pattern at the sample plane.
Meanwhile, the light outputted from the center tip could provide a uniform illumination. In other words, 4 raw images (three SI images and one wild-field image) can be captured by switching the channel of the optics switch.
As we previous proposed \cite{hu2019compact}, the frequency of the illumination patterns could be modulated by using different focal length combination of L1 and L2 or using different fiber mount with different distance between fiber tips.
It should be mentioned that L3 (125 mm) and L4 (180 mm) are build-in lenses of the microscope which is a 4F system, while L1 and L2 are not.
In practice, we places L1 and L2 as close as possible to get a higher power efficiency.

As we mentioned above, in our Method, we only captured one wide-field and three SI images, which did not contain any phase shift. Here we employed two methods to recover both SR and OS information from raw images.
As for SR part, a jRL deconvolution \cite{Strohl2015,Ingaramo2014} was used, which reformulated the matrix inversion problem as a minimization problem \cite{Str?hl2017}:
\begin{equation}
  S_{SR} = \min_{\hat{S}}[\sum_{n=1}^{N}(M_n - (I_n \times \hat{S}) \otimes h)^2], N \geq 3
  \label{eq: min}
\end{equation}
where $S_{SR}$ is the estimated high-resolution image of sample's fluorescence, $M_n$ is the $n^{th}$ captured image, $I_n$ is the $n^{th}$ illumination pattern, $\otimes$ is the convolution operator, $h$ is the point spread function (PSF).
This minimization problem can be solved based on iterations.
For each iteration in jRL deconvolution:
\begin{equation}
  \hat{M} = h \otimes (S_i \times I) + \mathcal{N},
  \label{eq: jrl1}
\end{equation}
\begin{equation}
  r = M/\hat{M},
  \label{eq: jrl2}
\end{equation}
\begin{equation}
  \hat{S}_{i+1} = \hat{S}_i \times [h \otimes (r \times I)]/[h \otimes(ones \times I)],
  \label{eq: jrl3}
\end{equation}
where $S_i$ is an estimate of the real fluorescence distribution of the sample in $i^{th}$ iteration, and $ones$ is a matrix of $1s$ with the same dimensions as the $M$.
By inputting all the captured images, $M_n$, and their illumination profile, $I_n$, iterations will yield high resolution images.
It should be mentioned that jRL deconvolution method has no OS capacity.
Fine details about the jRL deconvolution algorithm could be found in \cite{Strohl2015,Ingaramo2014}.
As for OS part, we employed a single-slice optical section algorithm based on Hilbert transform method \cite{hoffman2017single} to demodulate the in-focus information from each SI images.
Each structured illuminated image can be considered as the sum of modulated in-focus signal and unmodulated out-focus signal:
\begin{equation}
  M_l(r) = S_{OF_l}(r) + S_{IF_l}(r) \times cos (2 \pi k r + \phi),
  \label{eq: os1}
\end{equation}
where $S_{OF_l}(r)$ and $S_{IF_l}$ separately represents to the in- and out-focus components.
A notch filter (filtering out the $\pm 1$ peaks in Fourier domain) and a low-pass filter was used to estimate unmodulated component of each SI images, $R$.
Hence, the modulated in-focus component can be separated by subtracting $R$:
\begin{equation}
  M_{IF}(r) = M(r) - R(r).
  \label{eq: os2}
\end{equation}
Then, the in-focus image can be demodulated using Hilbert transform \cite{Ingaramo2014}:
\begin{equation}
  S_{IF} = \left| M_{IF} + i\mathcal{H} \right|.
  \label{eq: os3}
\end{equation}
Here $\mathcal{H}$ is the two-dimensional Hilbert transform of the $M_{IF}$,
\begin{equation}
  \mathcal{H} = \left| \mathcal{F}^{-1}[\mathcal{F}(M_{IF})\mathcal{S}] \right|,
\end{equation}
where $\mathcal{F}$ and $\mathcal{F}^{-1}$ represent to Fourier transform and inverse Fourier transform, and $\mathcal{S}$ is the spiral function in Fourier domain, $\mathcal{S}(u,v) = \frac{u+iv}{\sqrt{u^2+v^2}}$. Here $f$ is the spatial frequency.
By adopting this process to all the SI raw images and taking an average of them, the sectioned image, $S_{OS}$, could be calculated:
\begin{equation}
  S_{OS} = \frac{1}{N}\sum_{n=1}^{N}S_{IF_n}.
  \label{eq: os4}
\end{equation}
It should be noted that an accurate estimation of the unmodulated signal, $R$, is very important and also difficult during the demodulation using Hilbert transform \cite{Larkin2001}.

Finally, the high-resolution SR and the low-resolution OS information should be merged in to one final image.
We employed two strategies (\textbf{Strategy 1} and \textbf{Strategy 2}). The different performance of these two strategies will be discussed in more details in the experiment.

\textbf{Strategy 1}: We first estimate the SR image ($S_{SR}$) from measured raw images ($M_n$) using eq. \ref{eq: min}
and the OS image ($S_{OS}$) by using eq. \ref{eq: os4}.
Then we merged $S_{SR}$ and $S_{OS}$ images in Fourier domain \cite{kvrivzek2015simtoolbox,lukevs2014three} to get the final result ($S$):
\begin{equation}
  S = \mathcal{F}^{-1}[\beta\mathcal{F}(S_{OS} \mathrm{Mask}_{\alpha})+(1-\beta)\mathcal{F}(S_{SR})(1-\mathrm{Mask}_{\alpha})],
  \label{eq: Strategy 1}
\end{equation}
where $\mathrm{Mask}_{\alpha}$ is a low-pass filtering mask in Fourier domain, $\alpha$ represents to the radius of passband, and $\beta$ is a positive weight coefficient.

\textbf{Strategy 2}: We first calculate the in-focus information ($S_{IF_n}$) of each raw images ($M_n$) by using eq. \ref{eq: os3}.
Then the in-focus information ($M_{IF_n}$) could be estimated by subtracting their out-focus component:
\begin{equation}
  M_{IF_n} = M_n - \delta \mathcal{F}^{-1}[\mathcal{F}(M_4 - S_{IF_n}) \mathrm{Mask}_{\gamma}]
  \label{eq: Strategy 2}
\end{equation}
where $\mathrm{Mask}_{\gamma}$ is a low-pass filtering mask in Fourier domain, $\gamma$ represents to the radius of passband, $\delta$ is a positive weight coefficient, and $M_4$ is the wide-field image.
Final result ($S$) is reconstructed from $M_{IF_n}$ by using eq. \ref{eq: min}.

It should be mentioned that in the image processing we set the filter mask to be an ideal low-pass filter,

\begin{equation}
  \mathrm{Mask}_{\alpha} =
  \begin{cases}
    1 & u^2+v^2 \leq \alpha^2 \\
    0 & u^2+v^2 > \alpha^2
  \end{cases}
\end{equation}.

\section{Numerical Simulation}

\begin{figure}
  \centering
  \includegraphics{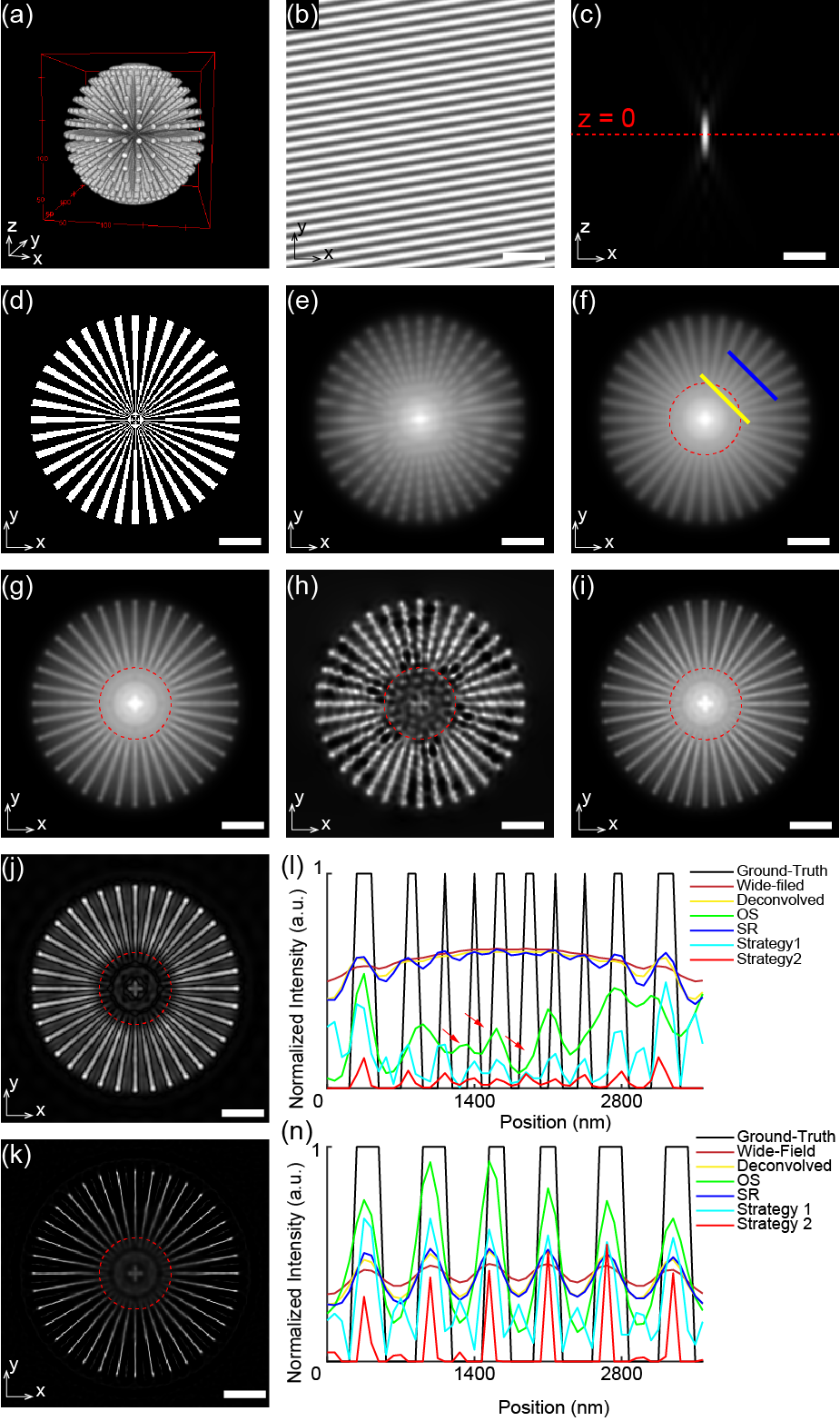}
  \caption{
  Simulation results.
  (a) a 3-dimention object with size of $256 \times 256 \times 256$ pixels;
  (b) one of the illumination patterns;
  (c) PSF using Born-and-Wolf model ($\mathrm{FWHM}_{xy} = 347.7$ nm; $\mathrm{FWHM}_z = 1140$ nm);
  (d) the ground-truth image of the object at z = 0 plane;
  (e) one of the raw SI images;
  (f) wide-field image;
  (g) deconvolved image;
  (h) deconvolved sectioned image using Hilbert transform method;
  (i) super-resolution image using jRL deconvolution method;
  (j) reconstructed image of Strategy 1;
  (k) reconstructed image of Strategy 2;
  (l) intensity profile along the yellow line shown in (f);
  (n) intensity profile along the blue line shown in (f).
  Here, we set $\alpha = 0.3 f_{cut-off}$, $\beta = 0.5$, $\gamma = 0.3 f_{cut-off}$ and $\delta = 1$.
  The red circle shown in the figure is corresponding to the resolution limit.
  The scale bars are in the figure are 2 $\mu m$.}
  \label{fig: simulation}
\end{figure}

To verify the performances of those methods we mentioned above, we used a 3-dimention object ($256 \times 256 \times 256$) shown in Figure \ref{fig: simulation} (a) to simulate the imaging process.
In the simulation,
the excited wavelength was 570 $nm$;
the numerical aperture of the imaging system was 1;
the size of each pixel was 50 $nm$.
Then the illumination pattern was set as
\begin{equation}
  I (r) = 0.5(1+cos(2 \pi \vec{k} r)),
\end{equation}
where $\vec{k}_n$ represents to the wave vector, and we set $|\vec{k}| = 0.6 f_{cut-off}$ which is close to the setting of our optical setup.
One of the illumination pattern is shown in Figure \ref{fig: simulation} (b).
The PSF we used in the simulation was generated using Born-and-Wolf model \cite{kirshner20133}, as shown in Figure \ref{fig: simulation} (c).
Mathematically, the acquirement of the raw SI images could be expressed as
\begin{equation}
  M(x,y,z_0) = \mathcal{P}((\sum^{Z}{(O(x,y,z)\times I(x,y))\otimes PSF(x,y,|z-z_0|)})_{\downarrow 2}),
\end{equation}
where $M(x,y,z_0)$ donates the measured image at $z=z_0$ plane, $O(x,y,z)$ represents to the 3D object,  $\downarrow 2$ donates a downsampling by a factor of 2, and $\mathcal{P}$ represents to the Poisson distribution.
In the simulation, one uniform-illuminated image (Figure \ref{fig: simulation}(f)) and three SI images in three orientations (Figure \ref{fig: simulation}(e)) at $z = 0$ plane were acquired.
Figure \ref{fig: simulation} (d) shows the ground-truth image of the object at z = 0 plane.

Figure \ref{fig: simulation}(g-k) shows the simulation results reconstructed by those methods we mentioned above:
(g) deconvolved image;
(h) deconvolved sectioned image using Hilbert transform method;
(i) super-resolution image using jRL deconvolution method;
(j) reconstructed image of Strategy 1;
(k) reconstructed image of Strategy 2.
In the reconstruction, we employed 50 jRL iterations in all the deconvolution processing, and we set $\alpha = 0.3 f_{cut-off}$, $\beta = 0.5$, $\gamma = 0.3 f_{cut-off}$ and $\delta = 1$.
The red circles in the figure are corresponding to the resolution limit.
Sub-diffraction details (within the red circle) were distinguished in Figure \ref{fig: simulation}(i-k), but not in Figure \ref{fig: simulation}(f-h), which shows that super-resolution results were achieved by using jRL deconvolution method.
As shown in Figure \ref{fig: simulation}(h,j,k), out-focus information was removed in the reconstructed results.
It should mentioned that the sectioned image, shown in Figure \ref{fig: simulation}(h), reconstructed by Hilbert transform method, contains some low-frequency artifacts.

In order to further compare the different super-resolution and optical sectioning performances of those reconstruction strategies, we analyzed two line intensity profiles (high-frequency and low-frequency area) pointed out in Figure \ref{fig: simulation}(f) by yellow and blue line.
As shown in Figure \ref{fig: simulation}(l) and (n), both \textbf{Strategy 1} and \textbf{2} could achieve good super-resolution and optical sectioning performances.
In high-frequency region (yellow line), the fine details in the result of \textbf{Strategy 1} were influenced by the low-frequency artifacts from the sectioned image (pointed out by red arrows), though they have a similar resolution comparing to the result of \textbf{Strategy 2}.
In relatively lower-frequency region (blue line), the result of \textbf{Strategy 2} has a clear background, but its structures are too thin comparing to the ground-truth image, which lost some information of the real object.

\begin{figure}
  \centering
  \includegraphics{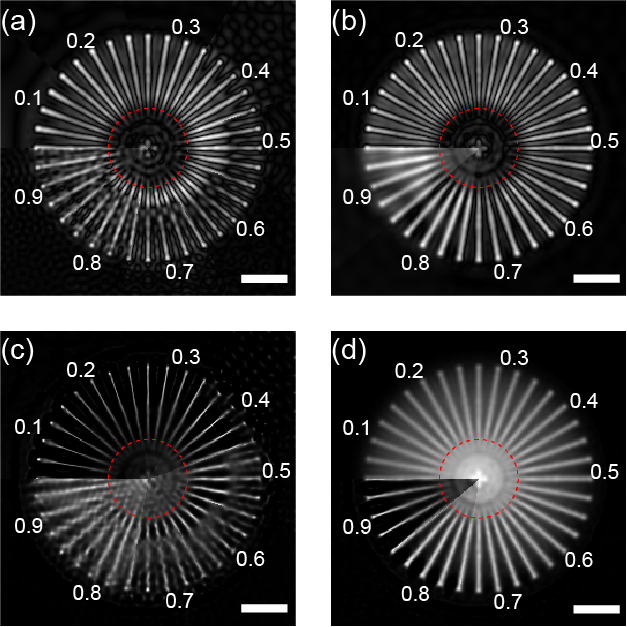}
  \caption{Reconstruction results using different settings of $\alpha$, $\beta$, $\gamma$ and $\delta$.
  (a) results of $\alpha = 0.1 \sim 0.9 f_{cut-off}$ and $\beta = 0.5$;
  (b) results of $\alpha = 0.3 f_{cut-off}$ and $\beta = 0.1 \sim 0.9$;
  (c) results of $\gamma = 0.1 \sim 0.9 f_{cut-off}$ and $\delta = 1$;
  (d) results of $\gamma = 0.3 f_{cut-off}$ and $\delta = 0.1 \sim 0.9$;
  The red circle shown in the figure is corresponding to the resolution limit.
  The scale bars are in the figure are 2 $\mu m$.
  }
  \label{fig: simulation_analysis}
\end{figure}

The settings of those variables' value in eq. \ref{eq: Strategy 1} ($\alpha$ and $\beta$) and eq. \ref{eq: Strategy 2} ($\gamma$ and $\delta$) make significant influence to the final imaging result; there is a trade-off in the combination of the super-resolution and optical sectioning.
Actually, these variables can be divided into two categories according to their functions: one is to match the spatial frequency range ($\alpha$ and $\gamma$); and another is to match the intensity ($\beta$ and $\delta$).
It should mentioned that, in the reconstruction, we assumed that all the out-focus information was low-frequency, which means that only the low-frequency part of the image needs to be sectioned.
In \textbf{Strategy 1}, the value of $\alpha$ ($0 \leq \alpha \leq f_{cut-off}$) determined the frequency range of the merging images; the low-frequency component ($f < \alpha$) of the OS image and the high-frequency component ($f > \alpha$) of the SR image are selected.
And these two components are finally added together by an intensity adjustment factor $\beta$ ($0 \leq \beta \leq 1$).
As shown in Figure \ref{fig: simulation_analysis} (a), if $\alpha$ is too small, there will be little optical sectioning effect; in contrast, is $\alpha$ is too large, the final result will contain the low-frequency artifacts produced by the single-slice optical section processing.
The influence of setting different value of $\beta$ is shown in Figure \ref{fig: simulation_analysis} (b).
Meanwhile, in \textbf{Strategy 2}, $\gamma$ ($0 \leq \gamma \leq f_{cut-off}$) plays a similar role as $\alpha$ in \textbf{Strategy 1}, which limited the frequency range of the out-focus component as shown in eq. \ref{eq: Strategy 2}.
And $\delta$ ($0 \leq \delta \leq 1$) is used to adjust the intensity of the out-focus information in order to obtain a good estimation of in-focus component of each raw images.
Those influences could be seen in Figure \ref{fig: simulation_analysis} (c) and (d).
However, the setting of these variables depends on the actual situation.

In all, \textbf{Strategy 2} that directly removes the out-focus components from the raw images provides better performance.

\section{Experiment}

In the experiment, we set the distance of each fiber tip pair to 6 mm, and focal length of L1 and L2 were 200 mm and 100 mm.
Considering the build-in lenses, L3 (125 mm) and L4 (180 mm), those point light sources (outputted by each fiber pairs) were projected on the back aperture of the OL (UPlanFLN, 60$\times$/1.25, Olympus) with a distance of 4.32 mm (the diameter of the back aperture of the OL was 7.5 mm), which means the SI patterns were set to $57.6\%$ of the maximum spatial frequency of the excitation light.
The numerical aperture of the OL can be adjusted from 0.65 to 1.25.
The light source used in the experiment was a 532 nm laser source with a maximum power of 100 mW, and a Cy3 filter cube ($\lambda_{ex} = 513 \sim 556 nm$ ; $\lambda_{em} = 570 \sim 613 nm$) was used to filter the light.

\begin{figure}
  \centering
  \includegraphics{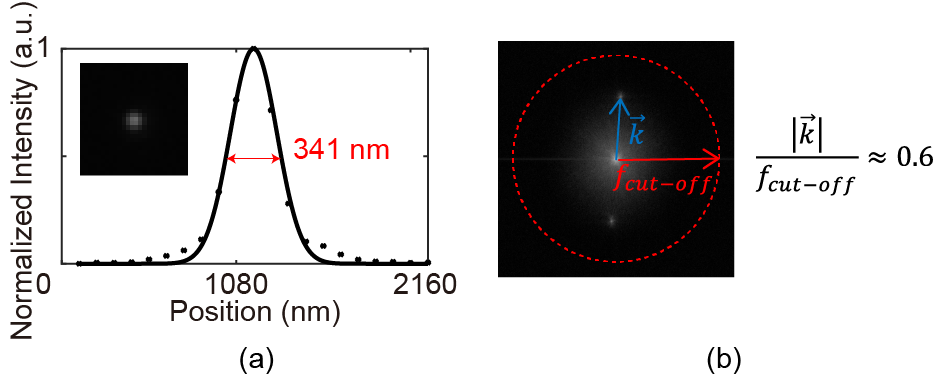}
  \caption{(a) shows the imaging result of a single 0.04 $\mu m$ fluorescent bead and the point spread funtion of our optical setup. (b) shows a frequency spectrum of a raw image with fringe illumination. The spatial frequency of the illumination pattern is about $0.6 f_{cut_off}$, which shows that the resolution enhancement factor of our setup is about 1.6.}
  \label{fig: rawdata}
\end{figure}

We first used 0.04 $\mu m$ fluorescent beads (F10720, Red, Thermo Fisher Scientific) to calibrate our microscope using the uniform illumination channel of the optics switch (ch4).
By averaging 22 separated point spots,  the point spread function (PSF) could be obtained by a first order Gaussian fitting, whose FWHM (full width at half maximum) is 341 nm in the experiment, as shown in Figure \ref{fig: rawdata}; the resolution limit of this imaging system is 341 nm.
As shown in Figure \ref{fig: rawdata}(b), The spatial frequency of the illumination pattern is about $0.6 f_{cut_off}$, which shows that the resolution enhancement factor of our setup is about 1.6 in theory.

\begin{figure}
  \centering
  \includegraphics{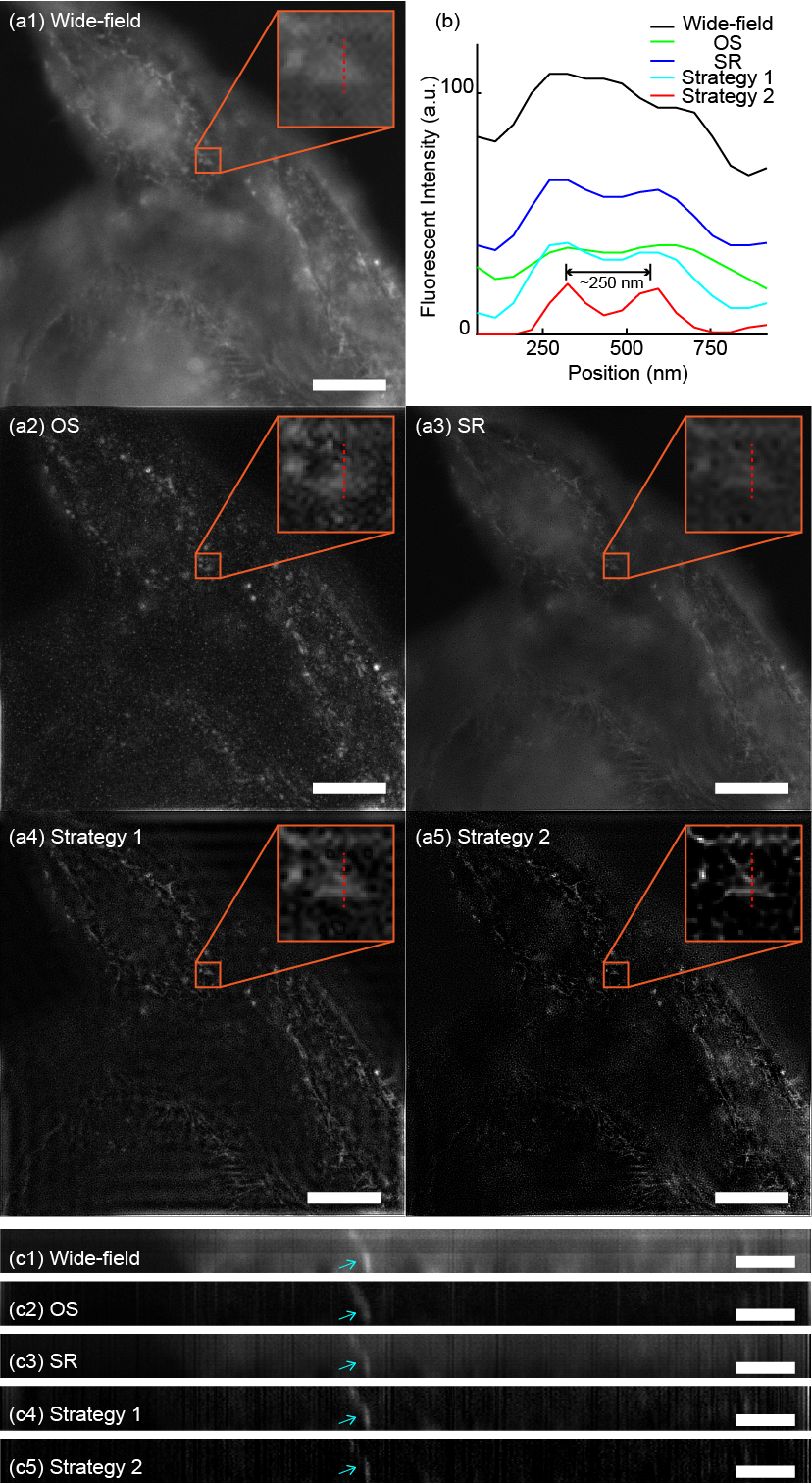}
  \caption{Fluorescent imaging results of Rhodamine Phalloidin labeled CaSki cells.
  We imaged 27 layers of the CaSki cells with a step of 0.1 $\mu m$ in Z-axis; 108 raw images were captured.
  A $1024\times1024$ pixel region of interest at $Z = 19$ layer was shown in (a).
  (a1): wide-field image;
  (a2): optical sectioned image by using Hilbert transform method;
  (a3): super-resolution image by using jRL deconvolution method;
  (a4): reconstructed image by using \textbf{Strategy 1};
  (a5): reconstructed image by using \textbf{Strategy 2};
  (b) shows a line fluorescent intensity distribution of the sample pointed out by red broken line.
  (c1)-(c5) show vertical cross-section images of the sample.
  The scale bars in (a) and (c) are $10 \mu m$ and $2 \mu m$}
  \label{fig: cell}
\end{figure}

Then we imaged Rhodamine Phalloidin ($\lambda_{ex} = 540 \sim 545 nm$ , $\lambda_{em} = 570 \sim 573 nm$) labeled CaSki cells to verify both of SR and OS performance of our method. 27 layers with a step of 0.1 $\mu m$ in vertical axis were imaged; four raw images were captured per layer.
The exposure time was set to 300 ms, and it took about 10 ms for the optics switch to switch illumination channel;
The acquisition time for each layer was about $1.230s$, and it took about $34.510s$ to complete the whole z-scanning.
Then we employed the image reconstruction methods described above, and both super-resolution and optical sectioned were achieved.
Fig. \ref{fig: cell} (a) shows a $1024\times1024$ pixels region.
A wide-field image is shown in Fig. \ref{fig: cell} (a1), while Fig. \ref{fig: cell} (a4) and (a5) show the reconstructed results by using \textbf{Strategy 1} and \textbf{2}.
To be more convincing, we also performed the Hilbert transform method and jRL deconvolution method individually, and the results are shown in Fig. \ref{fig: cell} (a2) and (a3).
In the image reconstruction, all jRL deconvolution processing was employed 50 iterations, and the variables were set as: $\alpha = 0.1f_{cut-off}$, $\beta = 0.5$, $\gamma = 0.1f_{cut-off}$ and $\delta = 1$, where $f_{cut-off}$ represents to the cut-off frequency of the microscope.
It takes about $36.157s$ for \textbf{Strategy 1} and $36.682s$ for \textbf{Strategy 2} to calculate one slice image ($1024 \times 1024$ pixels) by using a PC (Intel Core i7-6700 3.40GHz processor and 16GB RAM) with Matlab software (R2018b) under Windows 7 (SP1) x64 operating system.

As the line fluorescent distribution profile shown in Fig. \ref{fig: cell} (b), sub-diffraction resolution of $\sim 250 nm$ ($< 341 nm$) was achieved by using jRL deconvolution; the images reconstructed by employing only jRL deconvolution method, \textbf{Strategy 1} and \textbf{2} had been resolution enhanced.
To further verify the optical section performance, Fig. \ref{fig: cell} (c1) - (c5) show a vertical cross section images of the sample.
Low-resolution out-focus information was removed by using Hilbert transform method, but it lost lots of fine detail of the sample, as shown in Fig. \ref{fig: cell} (a2) and (c2).
Although fine detail was recover in Fig. \ref{fig: cell} (a3) and (c3), jRL deconvolution method cannot achieve optical sectioning.
Meanwhile, as shown in Fig. \ref{fig: cell} (c4) and (c5) pointed out by blue arrows, the structure is thin without any trailing in vertical axis, which indicated that both \textbf{Strategy 1} and \textbf{Strategy 2} could obtain super-resolution and optical sectioned images.
It also shows that more out-focus information was removed in Fig. \ref{fig: cell} (c5) than in (c4), while they have similar resolution.

\section{Discussion}

\begin{figure}
  \centering
  \includegraphics{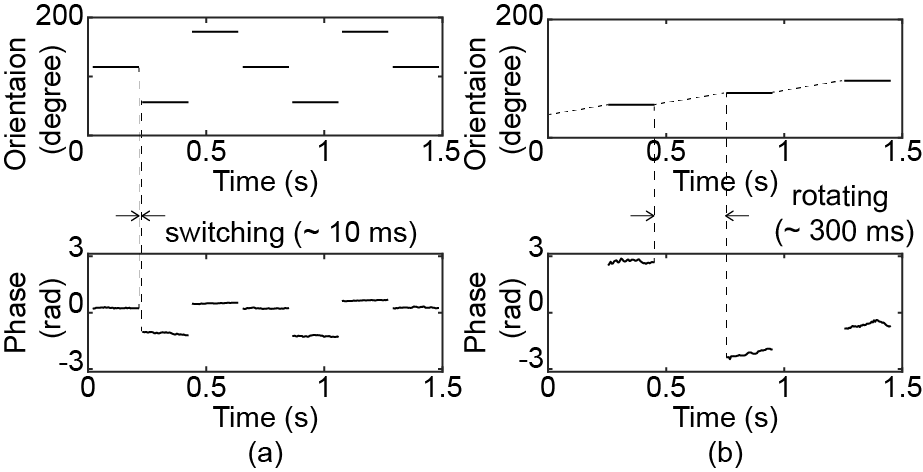}
  \caption{Stability analysis. We imaged some fluorescent dirt by setting the frame rate to 200 Hz. Both orientation and phase distribution of the SI patterns were shown.
  (a) is the results from optical switching; (b) is the results from mechanical rotating.}
  \label{fig: analysis}
\end{figure}

The main purpose of this work is to find a simple and low-cost way to build a stable SIM setup with acceptable imaging speed and similar imaging performance (both super-resolution and optical sectioning) comparing to other SIM setups.
For optical setup, there are two main parts of the SI pattern generation method.
One is the generation of two coherent point light sources.
We used three fiber coupler to split one laser source into point-light-sources pairs (the output tips of FCs), which could produce interference patterns at the sample plane in three orientations.
However, the single mode fused fiber optics coupler we used had a narrow band (532 $\pm$ 15 nm), which limited its application in multi-color imaging.
In our experience, a planar lightwave circuit (PLC) splitter with multiple operating wavelengths will be a preferred choice.
And another part is the switching of the illumination modes, which was achieved using a $1\times4$ optics switch whose switching time about $10 ms$; the max SI generation speed is about $100 Hz$.
Compared with our previous work \cite{hu2019compact}, the stability and the imaging speed was significantly improved in this approach.
As shown in Fig. \ref{fig: analysis}, we analyzed the different pattern modulation performance of both optics switch and mechanical rotation by imaging some fluorescent dirt from a highlighter pen.
The frame rate of the camera was set to 200 Hz, and the time period of each SI channels was set to 200 ms.
As shown, for mechanical rotation, there was a rapid phase shifting in random after rotation operation due to the inertia of the fibers in free space, which caused artifacts in image reconstruction, and it took about 300 ms to stabilize the pattern;
for optics switch, the phase was barely shifted in each orientations, and the orientation switching was fast, which only took about 10 ms.
Moreover, the phase was identical in each orientations by using optics switch, which reflected the stability improvement of the system in the other way.
However, the switching speed of the MEMS optics switch is not quite enough for ultra-high-speed fluorescent imaging.
Integrated optics devices seem to be a preferred solution of both small volume and high speed of channel switching in the future, and some related works have been already proposed recently \cite{helle2019structured,liu2016structured}.

Nowadays, SLMs are the most widely used devices to achieve accurate and high-speed structured illumination by loading different patterns on their chips.
Like grating, SLMs will diffract the incident laser light into multiple beams, which means they need a filter system to achieve two-beam of three-beam interference.
In addition, diffraction also causes low light efficiency.
Others devices like galvanometers or DOEs, they need to align multiple optical elements accurately.
Compared with those methods mentioned above, our method based on fiber devices made some improvements that by putting outputting fiber tips at the conjugate Fourier plane shorten the illumination path and also increases the light efficiency \cite{hu2019compact}.
Meanwhile, it is much simple and low-cost to build, while most of the fiber devices can be flexiblely assembled in small volume.
As a result, our method is more likely to be commercialized.

As for the image processing, 4-frame SIM data (three SI and one wide-field raw images) meets the minimum data acquirement to recover the real super-resolution image of the sample fluorescent distribution \cite{Str?hl2017}.
And it increased the frame rate in some ways, compared to other method based on 9- or 15-frame SI raw images.
Meanwhile, using a single-slice algorithm rather than other optical section algorithms based on $>1$ raw images \cite{neil1997method,zhou2015double} could avoid the influence of the uneven excitation intensity of different illumination channels, which could provide better sectioning performance.

In the combination of SR and OS, we employed two different strategies to combine the results of super-resolution and optical sectioning. Those two strategies have their own advantages and disadvantages.
In \textbf{Strategy 1}, the jRL deconvolution method and Hilbert transform method are performed separately, and then these two components were merged in Fourier domain.
It is suit for all kinds of SIM methods where the super-resolution and optical sectioning could individually performed.
However, the ``out-focus'' information will lower the modulation factor in pattern estimation.
In \textbf{Strategy 2}, the in-focus components of the raw SI images are demodulated by using Hilbert transform method first, and then we performed jRL deconvolution method with these ``in-focus'' data to reconstruct final result. It avoids to jRL deconvolve the out-focus component with illumination patterns.
However, it relies on a good estimation of uniform illuminated image, which makes it only suit for those SIM methods that could provide a wide-field image or yield a uniform-illuminated image by summing the SI images.

\section{Conclusion}
In summary, we proposed a compact and low-cost approach of generating structured illumination by using three fiber optics couplers and one MEMS optics switch.
By placing seven fiber tips at the conjugate Fourier plane of the illumination path, interference patterns were projected on the sample plane.
The switching of the illumination channels using the optics switch is stable and relatively high-speed.
Then super-resolution and sectioned image was reconstructed from 4 raw images by using algorithms combining jRL deconvolution method and Hilbert transform method.
Moreover, we shows that better imaging result could be obtained by directly removing the out-focus components from the raw images.
Our method provides a simple way to build a SIM setup under 4-frame SIM framework, which offers a similar imaging performance and acceptable imaging speed comparing to the conventional SIM setups and has its potential of commercialization in the future.

\section*{Acknowledgments}
This work was supported in part by the National Key Research and Development Program of China [2018YFF0214900], National Natural Science Foundation of China [21804126], and the Instrument Developing Project of the Chinese Academy of Sciences [YZ201402].

\section*{References}

\bibliography{references}

\begin{thebibliography}{10}
\expandafter\ifx\csname url\endcsname\relax
  \def\url#1{\texttt{#1}}\fi
\expandafter\ifx\csname urlprefix\endcsname\relax\def\urlprefix{URL }\fi
\expandafter\ifx\csname href\endcsname\relax
  \def\href#1#2{#2} \def\path#1{#1}\fi

\bibitem{Gustafsson2000}
M.~G.~L. Gustafsson, {Surpassing the lateral resolution limit by a factor of
  two using structured illumination microscopy.}, Journal of Microscopy 198~(2)
  (2000) 82--87.

\bibitem{neil1997method}
M.~A.~A. Neil, R.~Ju{\v{s}}kaitis, T.~Wilson, {Method of obtaining optical
  sectioning by using structured light in a conventional microscope}, Optics
  letters 22~(24) (1997) 1905--1907.

\bibitem{gustafsson2008three}
M.~G.~L. Gustafsson, L.~Shao, P.~M. Carlton, C.~J.~R. Wang, I.~N. Golubovskaya,
  W.~Z. Cande, D.~A. Agard, J.~W. Sedat, {Three-dimensional resolution doubling
  in wide-field fluorescence microscopy by structured illumination},
  Biophysical journal 94~(12) (2008) 4957--4970.

\bibitem{Wang2011}
L.~Wang, M.~C. Pitter, M.~G. Somekh, {Wide-field high-resolution structured
  illumination solid immersion fluorescence microscopy.}, Optics letters
  36~(15) (2011) 2794--2796.

\bibitem{Chang2009}
B.-J. Chang, L.-J. Chou, Y.-C. Chang, S.-Y. Chiang, {Isotropic image in
  structured illumination microscopy patterned with a spatial light modulator},
  Optics Express 17~(17) (2009) 14710.

\bibitem{dan2013dmd}
D.~Dan, M.~Lei, B.~Yao, W.~Wang, M.~Winterhalder, A.~Zumbusch, Y.~Qi, L.~Xia,
  S.~Yan, Y.~Yang, Others, {DMD-based LED-illumination super-resolution and
  optical sectioning microscopy}, Scientific reports 3 (2013) 1116.

\bibitem{liu2016structured}
Y.~Liu, C.~Wang, A.~Nemkova, S.-M. Hu, Z.-Y. Li, Y.-D. Yu, {Structured
  Illumination Chip Based on Integrated Optics}, Chinese Physics Letters 33~(5)
  (2016) 54204.

\bibitem{helle2019structured}
{\O}.~I. Helle, F.~T. Dullo, M.~Lahrberg, J.-C. Tinguely, B.~S. Ahluwalia,
  {Structured illumination microscopy using a photonic chip}, arXiv preprint
  arXiv:1903.05512.

\bibitem{liu2019three}
W.~Liu, Q.~Liu, Z.~Zhang, Y.~Han, C.~Kuang, L.~Xu, H.~Yang, X.~Liu,
  {Three-dimensional super-resolution imaging of live whole cells using
  galvanometer-based structured illumination microscopy}, Optics express 27~(5)
  (2019) 7237--7248.

\bibitem{rodriguez2008axial}
P.~F.~G. Rodriguez, E.~Sepulveda, B.~Dubertret, V.~Loriette, {Axial coding in
  full-field microscopy using three-dimensional structured illumination
  implemented with no moving parts}, Optics letters 33~(14) (2008) 1617--1619.

\bibitem{Wei2014}
F.~Wei, D.~Lu, H.~Shen, W.~Wan, J.~L. Ponsetto, E.~Huang, Z.~Liu, {Wide field
  super-resolution surface imaging through plasmonic structured illumination
  microscopy}, Nano Letters 14~(8) (2014) 4634--4639.

\bibitem{orieux2011bayesian}
F.~Orieux, E.~Sepulveda, V.~Loriette, B.~Dubertret, J.-C. Olivo-Marin,
  {Bayesian estimation for optimized structured illumination microscopy}, IEEE
  Transactions on image processing 21~(2) (2011) 601--614.

\bibitem{Dong2015}
S.~Dong, J.~Liao, K.~Guo, L.~Bian, J.~Suo, G.~Zheng, {Resolution doubling with
  a reduced number of image acquisitions}, Biomedical Optics Express 6~(8)
  (2015) 2946--2952.

\bibitem{Lal2016a}
A.~Lal, X.~Huang, P.~Xi, {A frequency domain reconstruction of SIM image using
  four raw images}, in: Optics in Health Care and Biomedical Optics VII, Vol.
  10024, International Society for Optics and Photonics, 2016, p. 1002411.

\bibitem{Meiniel2017}
W.~Meiniel, P.~Spinicelli, E.~D. Angelini, A.~Fragola, V.~Loriette, F.~Orieux,
  E.~Sepulveda, J.-C. Olivo-Marin, {Reducing data acquisition for fast
  Structured Illumination Microscopy using Compressed Sensing}, in: Biomedical
  Imaging (ISBI 2017), 2017 IEEE 14th International Symposium on, IEEE, 2017,
  pp. 32--35.

\bibitem{hoffman2017single}
Z.~R. Hoffman, C.~A. DiMarzio, {Single-image structured illumination using
  Hilbert transform demodulation}, Journal of biomedical optics 22~(5) (2017)
  56011.

\bibitem{jost2015optical}
A.~Jost, E.~Tolstik, P.~Feldmann, K.~Wicker, A.~Sentenac, R.~Heintzmann,
  {Optical sectioning and high resolution in single-slice structured
  illumination microscopy by thick slice blind-SIM reconstruction}, PloS one
  10~(7) (2015) e0132174.

\bibitem{soubies2018computational}
E.~Soubies, M.~Unser, {Computational Super-Sectioning for Single-Slice
  Structured-Illumination Microscopy}, IEEE Transactions on Computational
  Imaging.

\bibitem{hu2019compact}
S.~Hu, L.~Liu, Y.~Huang, W.~Liu, Q.~Wei, M.~Tan, Y.~Yu, {Compact and low-cost
  structured illumination microscopy using an optical fiber coupler}, Optics
  Communications 436 (2019) 227--231.

\bibitem{Strohl2015}
F.~Str{\"{o}}hl, C.~F. Kaminski, {A joint Richardson—Lucy deconvolution
  algorithm for the reconstruction of multifocal structured illumination
  microscopy data}, Methods and Applications in Fluorescence 3~(1) (2015)
  014002.

\bibitem{Ingaramo2014}
M.~Ingaramo, A.~G. York, E.~Hoogendoorn, M.~Postma, H.~Shroff, G.~H. Patterson,
  {Richardson-Lucy deconvolution as a general tool for combining images with
  complementary strengths}, ChemPhysChem 15~(4) (2014) 794--800.
\newblock \href {http://arxiv.org/abs/NIHMS150003} {\path{arXiv:NIHMS150003}}.

\bibitem{Str?hl2017}
F.~Str{\"{o}}hl, C.~F. Kaminski, {Speed limits of structured illumination
  microscopy}, Optics Letters 42~(13) (2017) 2511.

\bibitem{Larkin2001}
K.~G. Larkin, D.~J. Bone, M.~A. Oldfield, {Natural demodulation of
  two-dimensional fringe patterns I General background of the spiral phase
  quadrature transform}, Journal of the Optical Society of America A 18~(8)
  (2001) 1862.

\bibitem{kvrivzek2015simtoolbox}
P.~Krizek, T.~Luke{\v{s}}, M.~Ovesn{\'{y}}, K.~Fliegel, G.~M. Hagen,
  {SIMToolbox: a MATLAB toolbox for structured illumination fluorescence
  microscopy}, Bioinformatics 32~(2) (2015) 318--320.

\bibitem{lukevs2014three}
T.~Luke{\v{s}}, P.~Kř{\'{i}}{\v{z}}ek, Z.~{\v{S}}vindrych, J.~Benda,
  M.~Ovesn{\'{y}}, K.~Fliegel, M.~Kl{\'{i}}ma, G.~M. Hagen, {Three-dimensional
  super-resolution structured illumination microscopy with maximum a posteriori
  probability image estimation}, Optics express 22~(24) (2014) 29805--29817.

\bibitem{kirshner20133}
H.~Kirshner, F.~Aguet, D.~Sage, M.~Unser, 3-d psf fitting for fluorescence
  microscopy: implementation and localization application, Journal of
  microscopy 249~(1) (2013) 13--25.

\bibitem{zhou2015double}
X.~Zhou, M.~Lei, D.~Dan, B.~Yao, J.~Qian, S.~Yan, Y.~Yang, J.~Min, T.~Peng,
  T.~Ye, Others, {Double-exposure optical sectioning structured illumination
  microscopy based on Hilbert transform reconstruction}, PloS one 10~(3) (2015)
  e0120892.

\end{thebibliography}

\end{document}